\documentclass[onecolumn,showpacs,preprintnumbers,amsmath,amssymb]{revtex4-1}
\usepackage{graphicx}% Include figure files
\usepackage{dcolumn}% Align table columns on decimal point
\usepackage{bm}% bold math

\usepackage{color}
\usepackage{ulem}

\begin{document}

\preprint{}

\title{$R$-parity violating supersymmetric contributions \\
to the neutron beta decay at the one-loop level}% Force line breaks with \\

\author{Nodoka Yamanaka}
\affiliation{%
Research Center for Nuclear Physics, Osaka University, Ibaraki, Osaka 567-0047, Japan}%

\author{Toru Sato and Takahiro Kubota}
\affiliation{%
Department of Physics, Osaka University, Toyonaka, Osaka 560-0043, Japan}%

\date{\today}% It is always \today, today,

\begin{abstract}
The contribution of the $R$-parity violating minimal supersymmetric standard model to the neutron beta decay at the one-loop level is investigated.
It is found that the baryon number and $R$-parity violating interactions contribute to the $D$ correlation through one-loop corrections, while the tree-level prediction is vanishing. 
The Fierz interference term is also investigated at the one-loop level by considering the lepton number and $R$-parity violating interactions.
We show that future experimental progress can provide us with better constraints on some of the combinations of $R$-parity violating couplings.

\end{abstract}

\maketitle

\section{\label{sec:intro}Introduction}
The standard model (SM) of particle physics is known to be very successful in interpreting many experimental data up to now. 
There are however some phenomena which are difficult to explain in this framework, such as the matter abundance of our Universe, the absence of candidates of dark matter, etc. 
We need therefore to introduce some new physics (NP) beyond the SM.

One approach to search for NP is the fundamental test of low energy phenomena, which consists of the precision measurement of experimental observables with well known SM predictions. 
By observing discrepancies from the SM data, we can establish the existence of NP.
One interesting phenomenon which can probe the NP is the beta decay of the neutron and nuclei \cite{herczeg2,erler,severijns,abele,hewett}. 
The beta decay provides many observables \cite{jackson} sensitive to NP, such as the Fierz interference term \cite{hardy}, the $D$ correlation \cite{halin,soldner,mumm}, the $R$ correlation \cite{schneider,sromicki,kozela}, etc, and their experimental developments in recent years are very promising.
As these observables have very small SM predictions \cite{herczeg1}, we can say that they are an excellent probe of NP.

On the theoretical side, the minimal supersymmetric standard model (MSSM) \cite{mssm} is known to be one of the leading candidates of the NP. 
A general supersymmetric extension of the SM allows baryon number or lepton  number violating interactions, so we must impose the conservation of {\it R-parity} ($R=(-1)^{3B-L +2s}$) to forbid them. 
This assumption is however completely {\it ad hoc}, so the $R$-parity violating (RPV) interactions need to be investigated phenomenologically. 
Until now, many of the RPV interactions were constrained by high energy experiments, low energy precision tests, and cosmological phenomenology \cite{chemtob, barbier,rpvphenomenology}.

In the ($R$-parity conserving) MSSM, the contribution to the $D$ correlation of neutron beta decay was found to be at most on the order of $10^{-7}$ \cite{christova,drees}. 
In the RPV sector, the discussion is divided into two distinct cases where either baryon or lepton number violating interactions are involved but not both, since their coexistence is strongly forbidden by the nonobservation of the proton decay.
The separate analyses of the effects of baryon and lepton number violating RPV interactions to the beta decay must therefore be performed.
The lepton number violating RPV interactions contribute to the beta decay at the tree-level, and it was found that the Fierz interference term and the $R$ correlation are sensitive observables to RPV interactions \cite{barger,herczeg2,herczeg3,kao,yamanaka1}.

The alternative case, the baryon number violating $R$-parity violation, generates the $D$ correlation starting from the one-loop level.
This was analyzed in Ref. \cite{ng}, yielding new constraints on some combinations of RPV couplings using the relation between the $D$ correlation and the electric dipole moment (EDM) of the neutron.
This previous loop level analysis of the RPV sector, however, did not cover all of the one-loop diagrams, and as we will show,
there exist additional contributions to the $D$ correlation.

For the lepton number violating RPV interactions, there are also new contributions which appear  at the one-loop level through flavor change, and generate the Fierz interference term as an observable effect.
These new contributions involve  different combinations of RPV couplings, and their sparticle mass dependencies also differ from the tree-level.
It could be that the one-loop level effect surpasses the tree-level one.
We have therefore good reasons to discuss the one-loop contribution.
In this case we have found that similar techniques used in the analysis of the one-loop level P, CP-odd electron-nucleon interaction c-an be applied \cite{yamanaka2}.

Now that we have sufficient motivations, we will analyze in this paper the complete set of RPV contributions to the beta decay at the one-loop level and give its potential observable signature.
Our discussion is organized as follows.
We first briefly review the RPV interactions in the next section.
In Section \ref{sec:rpv_quark_decay}, we classify and give the RPV contributions to the quark beta decay at the one-loop level.
In Section \ref{sec:obs}, we relate the quark level physics to the neutron beta decay effective interactions by introducing the relevant nucleon matrix elements and present the observables in question, the $D$ correlation and the Fierz interference term of the neutron beta decay.
In Section \ref{sec:dcorr}, we analyze the baryon number violating RPV contribution to the $D$ correlation of the beta decay and give the constraints on RPV couplings which can be provided by future beta decay experiments.
We also give a short comment on the constraints given by other experiments, in particular those provided by the EDM.
In Section \ref{sec:fierz}, the analysis of the lepton number violating RPV contribution to the Fierz interference term is presented.
The final section is devoted to the summary.

\section{\label{sec:RPV}RPV lagrangian}

Let us first introduce the RPV interactions.
The superpotential of the RPV interactions can be written as follows:
\begin{eqnarray}
W_{R\hspace{-.5em}/} &=& \frac{1}{2} \lambda_{ijk} \epsilon_{ab}
 L_i^a L_j^b (E^c)_k
 +\lambda'_{ijk} \epsilon_{ab} L_i^a Q_j^b ( D^c)_k \nonumber\\
&& + \frac{1}{2} \epsilon_{RGB} \lambda''_{ijk} (U^c)_i^R (D^c)_j^G (D^c)_k^B \ ,
\label{eq:superpotential}
\end{eqnarray}
with $i,j,k=1,2,3$ indicating the generation, $a,b=1,2$ the $SU(2)_L$, and $R,G,B = 1,2,3$ the color $SU(3)_c$ indices, respectively.
The lepton left-chiral superfields $L$ and $E^c$ are respectively $SU(2)_L$ doublet and singlet. 
The quark superfields $Q$, $U^c$ and $D^c$ denote respectively the quark $SU(2)_L$ doublet, up-quark singlet and down-quark singlet left-chiral superfields.
The bilinear term has been omitted in our discussion. 
We have also neglected the soft breaking terms in the RPV sector.
We should note that the coexistence of lepton number violating interactions ($\lambda_{ijk}$ and $\lambda'_{ijk}$) and baryon number violating interactions ($\lambda''_{ijk}$) induces rapid proton decay \cite{smirnov}, so we must investigate them separately.
The above RPV superpotential gives the following lepton and baryon number violating Yukawa interactions:
\begin{widetext}
\begin{eqnarray}
{\cal L }_{L\hspace{-.4em}/\,} &=&
- \frac{1}{2} \lambda_{ijk} \left[
\tilde \nu_i \bar e_k P_L e_j +\tilde e_{Lj} \bar e_k P_L \nu_i + \tilde e_{Rk}^\dagger \bar \nu_i^c P_L e_j -(i \leftrightarrow j ) \right] + ({\rm h.c.})\nonumber\\
&&-\lambda'_{ijk} \left[
\tilde \nu_i \bar d_k P_L d_j + \tilde d_{Lj} \bar d_k P_L \nu_i +\tilde d_{Rk}^\dagger \bar \nu_i^c P_L d_j -\tilde e_{Li} \bar d_k P_L u_j - \tilde u_{Lj} \bar d_k P_L e_i - \tilde d_{Rk}^\dagger \bar e_i^c P_L u_j \right]  + ({\rm h.c.}) \ ,\nonumber\\
{\cal L }_{B\hspace{-.5em}/\,}&=&- \frac{1}{2} \lambda''_{ijk} \epsilon_{RGB} \left[
\tilde d_{Rk}^{B\dagger} \bar u_i^R P_L d^{cG}_j +\tilde d_{Rj}^{G\dagger} \bar u_i^R P_L d^{cB}_k +\tilde u_{Ri}^{R\dagger} \bar d_j^G P_L d^{cB}_k -(j \leftrightarrow k )
\right] + ({\rm h.c.}) \ .
\end{eqnarray}
\end{widetext}
The matter fields of the above lagrangian are assumed to be mass eigenstates.

\section{\label{sec:rpv_quark_decay}RPV contribution to the quark beta decay at one-loop level}
\begin{figure}[htb]
\includegraphics[width=7.2cm]{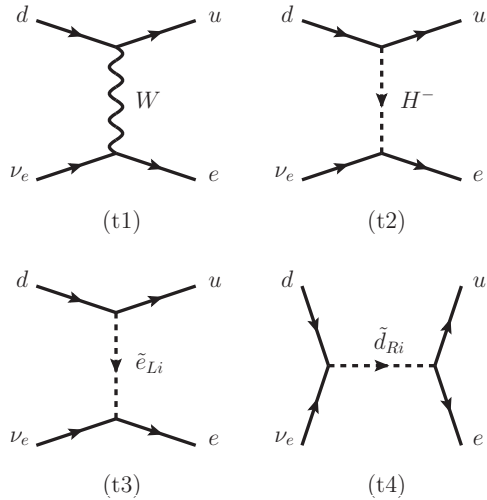}
\caption{\label{fig:rpv_tree} Tree-level contribution to the quark beta decay within RPVMSSM.
Diagrams (t3) and (t4) are generated by RPV interactions.
}
\end{figure}

\begin{figure}[htb]
\includegraphics[width=14cm]{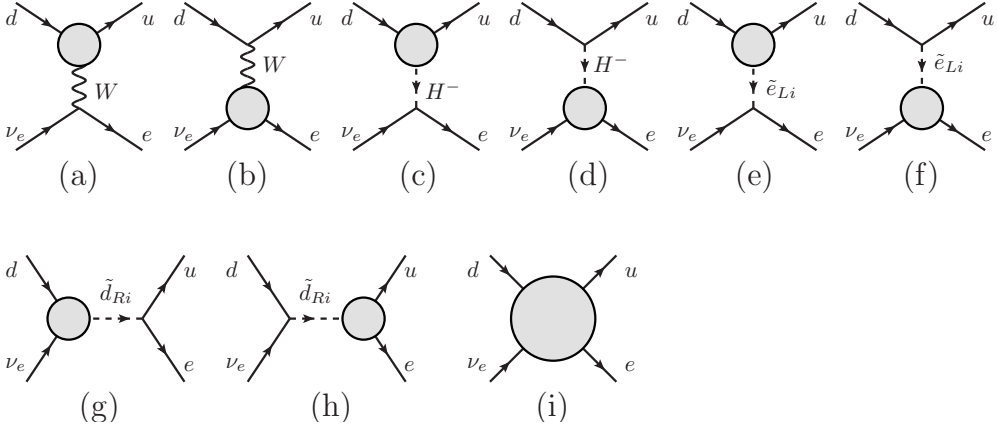}
\caption{\label{fig:classification} Classification of one-loop correction to the beta decay amplitude within RPVMSSM.
The grey blobs denote the one-loop effective vertex.
}
\end{figure}

Let us now show the RPV contribution to the beta decay at one-loop level.
On the basis of tree-level contributions (see Fig. \ref{fig:rpv_tree}), we can classify the RPV corrections contributing to the quark beta decay at the  one-loop level as shown in Fig. \ref{fig:classification}.
Let us see them one by one in detail. (Incidentally, tree diagrams in Fig. \ref{fig:rpv_tree} were analyzed in Refs. \cite{herczeg2,yamanaka1}).\\

\begin{figure}[htb]
\includegraphics[width=8cm]{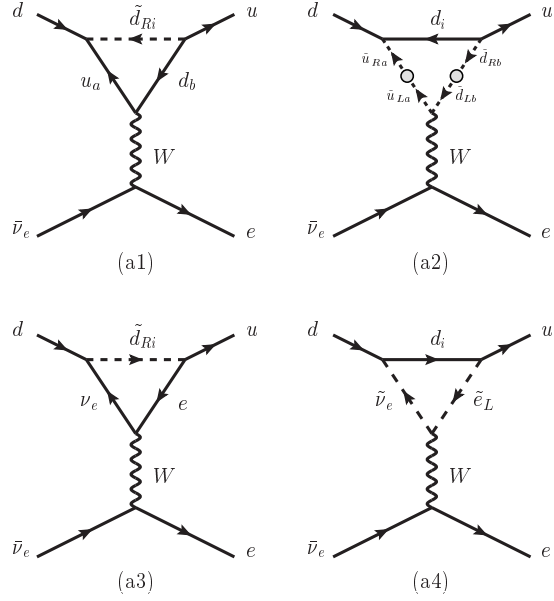}
\caption{\label{fig:loop_wqq} 
R-parity violating correction to the $W$ boson-quark vertex at the one-loop level.
Diagrams (a1) and (a2) contribute to the $(V+A)\times (V-A)$ interaction, while (a3) and (a4) are the $(V-A)\times (V-A)$ interaction.
The grey blobs denote mass insertions which mix the squark gauge eigenstates.
}
\end{figure}

{\it Wqq corrections} (diagram (a) of Fig. \ref{fig:classification}):\\
This is the RPV correction to the SM contribution with $W$ boson exchange (Fig. \ref{fig:rpv_tree} (t1) ).
The complete set of this type is depicted in Fig. \ref{fig:loop_wqq}.
The amplitude of the first diagram (a1) can be written as
\begin{eqnarray}
i{\cal M}_{\rm (a1)} & = & 2 i \frac{ \lambda''_{1bi} \lambda''^*_{a1i} V_{ab} m_{u_a} m_{d_b} }{(4\pi)^2 } \frac{G_F}{\sqrt{2}} I(m_{u_a}^2 , m_{d_b}^2 , m_{\tilde d_{Ri}}^2)  \bar u \gamma_\mu (1+\gamma_5) d \, \bar e \gamma^\mu (1-\gamma_5) \nu_e \ , 
\label{eq:(a1)}
\end{eqnarray}
where we have neglected the external and exchanged momenta. 
Indices $i,a$ and $b$ indicate the flavor.
Here $G_F$ is the Fermi constant.
The CKM matrix is denoted by $V_{ab}$.
The loop integral $I$ is expressed as follows:
\begin{eqnarray}
I(a,b,c) 
&=&
\frac{1}{(b-a)(c-b)(a-c)} \left[ ab \ln \frac{a}{b} +bc \ln \frac{b}{c} +ca \ln \frac{c}{a} \right] .
\label{eq:idef}
\end{eqnarray}
For $m_{\tilde d_{Ri}}^2 >> m_{u_a}^2 , m_{d_b}^2$, we have
\begin{eqnarray}
I(m_{u_a}^2 , m_{d_b}^2 , m_{\tilde d_{Ri}}^2) 
&\approx & \frac{1}{(m_{u_a}^2 - m_{d_b}^2 ) m_{\tilde d_{Ri}}^2 } \left[ m_{u_a}^2 \ln \frac{m_{\tilde d_{Ri}}^2}{m_{u_a}^2}- m_{d_b}^2 \ln \frac{m_{\tilde d_{Ri}}^2}{m_{d_b}^2}\right] \ .\nonumber\\ 
\end{eqnarray}
We see from eq. (\ref{eq:(a1)}) that the process (a1) is a $(V+A)\times (V-A)$ interaction.
As the product of RPV interaction $\lambda''_{1bi} \lambda''^*_{a1i} $ can have complex phase, the imaginary part of this amplitude contributes to the CP-odd $(V+A)\times (V-A)$ interaction.
As observable sensitive to the CP-odd $(V+A)\times (V-A)$ interaction, we have the $D$ correlation of the beta decay.
The contribution of eq. (\ref{eq:(a1)}) was not discussed in Ref. \cite{ng}.
The RPV one-loop level contribution to the $D$ correlation is discussed in Sections \ref{sec:obs} and \ref{sec:dcorr}.

The diagram (a2) is the analogue of (a1) with all fields in the loop interchanged with their superpartner, and involves exactly the same RPV couplings as the (a1) contribution.
This contribution was treated in Ref. \cite{ng}.
It is also of $(V+A)\times (V-A)$ type interaction.
The amplitude of the second diagram (a2) is given by
\begin{eqnarray}
i{\cal M}_{\rm (a2)} 
& = & 
-i \frac{ \lambda''_{1bi} \lambda''^*_{a1i} V_{ab} C_{u_a} C_{d_b}}{(4\pi)^2 } \frac{G_F}{\sqrt{2}} 
J (m_{\tilde u_L}^2 , m_{\tilde u_R}^2 , m_{\tilde d_L}^2, m_{\tilde d_R}^2) 
 \bar u \gamma_\mu (1+\gamma_5 ) d \, \bar e \gamma^\mu (1-\gamma_5 ) \nu_e ,
\label{eq:(a2)}
\end{eqnarray}
where $C_{u_a}$ and $C_{d_b}$ are mass insertions given as follows:
\begin{eqnarray}
C_{u_a} &=& -m_{u_a} (-A_{u_a} +\mu \cot \beta )\ ,\\
C_{d_b} &=& -m_{d_b} (-A_{d_b} +\mu \tan \beta )\ .
\end{eqnarray}
If we take $\tan \beta =50$ and soft parameters $A$ and $\mu$ around 1 TeV, $C_{u_a}C_{d_b} \approx -m_{u_a}m_{d_b}A_{u_a} \mu \tan \beta$.
The loop integral $J$ is expressed as:
\begin{equation}
J(a,b,c,d) \equiv \int_0^\infty \frac{r dr}{(r+a)(r+b)(r+c)(r+d)}.
\end{equation}
It is not profitable to examine every corner of the parameter space of $m_{\tilde u_L}^2 , m_{\tilde u_R}^2 , m_{\tilde d_L}^2$ and $m_{\tilde d_R}^2$.
Here we take them to be equal in magnitude.
We have then
\begin{equation}
J(m_{\rm SUSY}^2,m_{\rm SUSY}^2,m_{\rm SUSY}^2,m_{\rm SUSY}^2) = \frac{1}{6 m_{\rm SUSY}^4}.
\end{equation}
We must note that signs of soft parameters $A$ and $\mu$ are so far undetermined, so there is a possibility of cancellation between contributions (a1) and (a2).

Let us add a brief comment on the Lorentz structure of (a1) and (a2).
At first sight, it may look strange to obtain a right-handed vector current from $W$ boson interacting vertex.
In the case in question, this was possible thanks to the right-chirality projection of the external down quark due to baryon number violating RPV interactions ($\lambda''$) and also to the propagation of antiparticles in the loop.
We will see that for other diagrams, this is not possible.

The remaining (a3) and (a4) diagrams of Fig. \ref{fig:loop_wqq} contribute to the $(V-A)\times (V-A)$ interaction of the beta decay.
The tree-level SM contribution has also the same $(V-A)\times (V-A)$ structure, so these remaining diagrams work as a shift of the Fermi constant $G_F$.
As the redefinition of the Fermi constant cannot be probed by the nucleon beta decay itself, we do not consider them.
In the case of (a3) and (a4), the projection of the chirality of external down quark is left-handed, so it is not possible to generate $(V+A)$ quark current.
This is due to the chirality structure of the lepton number violating RPV interactions ($\lambda$ and $\lambda'$).
\\

\begin{figure}[htb]
\includegraphics[width=8cm]{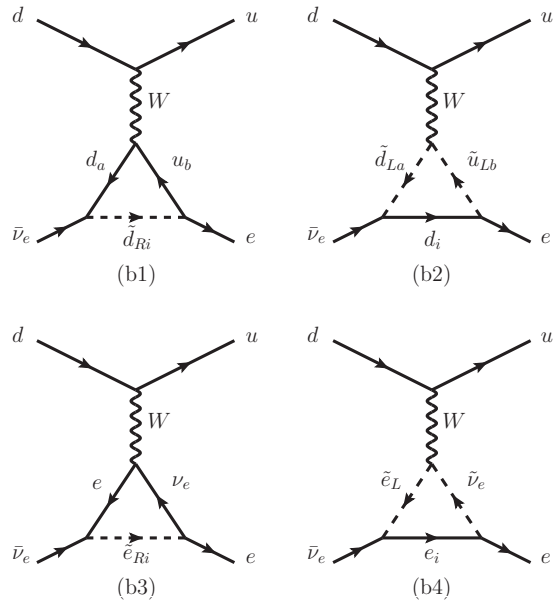}
\caption{\label{fig:loop_wll}
R-parity violating correction to the $W$ boson-lepton vertex at the one-loop level.
All diagrams have the $(V-A)\times (V-A)$ structure.}
\end{figure}

{\it Wll corrections} (diagram (b) of Fig. \ref{fig:classification}):\\
As the $Wqq$ corrections, there are also RPV corrections to the $W$ boson-lepton vertex.
The complete set of this type is depicted in Fig. \ref{fig:loop_wll}.
For the case of $Wll$ corrections, all diagrams have the $(V-A)\times (V-A)$ structure.
This is because the chirality projection of the neutrino due to lepton number violating RPV interactions ($\lambda$ and $\lambda'$) gives only left-handed lepton currents.
As we have seen, the $(V-A)\times (V-A)$ interaction gives the shift of the Fermi constant, so they do not lead to an observable effect.
We therefore do not consider the $Wll$ corrections in this discussion.
\\

{\it Corrections to charged Higgs exchange} (diagrams (c) and (d) of Fig. \ref{fig:classification}):\\
The Higgs exchange contribution to the beta decay can be drawn by replacing the $W$ boson of diagrams (a) and (b) by the charged Higgs boson.
These radiative corrections give a scalar-type interaction of beta decay.
They are however suppressed by at least a factor of light fermion Yukawa coupling (smaller than $O(10^{-5})$), so their effects are negligible.
We do not consider them further in our discussion.
\\

{\it $\tilde e_L$-fermion vertex corrections} (diagrams (e) and (f) of Fig. \ref{fig:classification}):\\
These one-loop corrections are the vertex corrections to the tree-level selectron exchange diagram (Fig. \ref{fig:rpv_tree} (t3)).
As the vertex corrections are the renormalization of the RPV interactions, we do not need to consider this set of diagrams.
\\

{\it $\tilde d_R$-fermion vertex corrections} (diagrams (g) and (h) of Fig. \ref{fig:classification}):\\
This type corresponds to the vertex corrections to the tree-level down-squark exchange contribution (Fig. \ref{fig:rpv_tree} (t4)).
Again, they do not need to be treated as the vertex corrections are renormalization of the tree-level RPV interactions.
\\

{\it Box diagrams} (diagram (i) of Fig. \ref{fig:classification}):\\
The one-loop level box diagrams yield finite contributions to the beta decay amplitude.
They correspond to the electroweak radiative corrections to the tree-level RPV processes (Fig. \ref{fig:rpv_tree} (t3) and (t4)).
The QCD radiative corrections to the RPV amplitude are not considered, since the hadron matrix elements given in the next section include them nonperturbatively.
Corrections with Higgs bosons are neglected, since their contributions receive suppression from light fermion Yukawa couplings.

The box diagrams can be classified into two types.
The first type is the flavor conserving contribution, given by photon, $Z$ boson and neutralino corrections.
After diagrammatic analysis, we have found that these one-loop diagrams give only higher-order corrections in $\alpha_{\rm em}$ to the tree-level selectron or down-squark exchange contributions (Fig. \ref{fig:rpv_tree} (t3) and (t4)), with the same Lorentz structure (scalar, pseudoscalar interactions, see Appendix \ref{sec:appendix} for detail).
We cannot expect them to yield particular observable effects, until RPV interactions are discovered and quantitatively studied at the hig-her order in $\alpha_{\rm em}$.
It is then not useful at present to treat them.

\begin{figure}[htb]
\includegraphics[width=8cm]{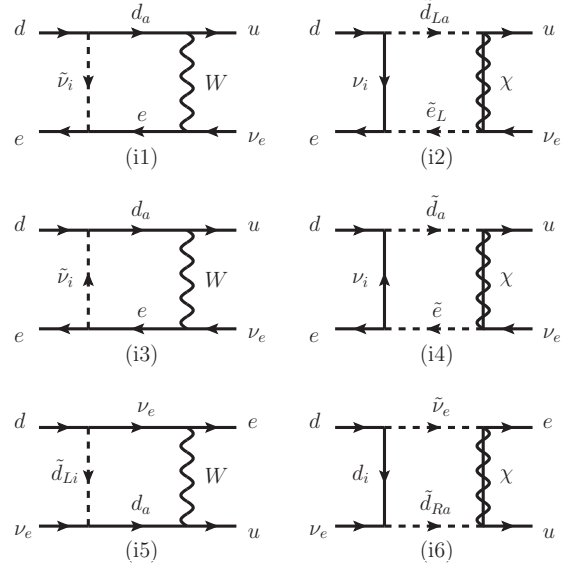}
\caption{\label{fig:fcbox}
Box diagrams with flavor change contributing to the beta decay at the one-loop level in RPVMSSM.
The chargino is denoted by $\chi$.
}
\end{figure}

The second type is the flavor changing contribution, given by $W$ boson and chargino corrections.
This contribution, although being suppressed against the tree-level ones, involves RPV flavor structures not relevant at the tree-level, and is thus interesting.
The relevant diagrams are shown in Fig. \ref{fig:fcbox}

The contribution of the first diagram (i1) is
\begin{eqnarray}
i{\cal M}_{\rm (i1)}&=&
i \frac{\lambda_{i11} \lambda'^*_{ia1}e^2 V_{1a}}{8(4\pi)^2 \sin^2 \theta_W} I(m_{d_a}^2 , m_{\tilde \nu_i}^2 , m_W^2 ) 
\bar u (1+\gamma_5) d \, \bar e (1-\gamma_5) \nu_e .
\label{eq:(i1)}
\end{eqnarray}
The loop integral $I$ was defined in eq. (\ref{eq:idef}).
As we can see, this amplitude yields scalar and pseudoscalar interactions.
The scalar-type interaction contributes to the Fierz interference term (CP-even part) and  to the $R$ correlation (CP-odd part) of the beta decay.
The pseudoscalar interaction vanishes in the nonrelativistic limit of the nucleon, so we neglect it from now.
The RPV one-loop level contribution to these observables is discussed in the next section.

The second diagram (i2) is the analogue of (i1), with all particles in the loop interchanged by their superpartner.
It is expressed as
\begin{eqnarray}
i{\cal M}_{\rm (i2)}&=&
i \frac{\lambda_{i11} \lambda'^*_{ia1}e^2 V_{1a}}{8(4\pi)^2 \sin^2 \theta_W} \sum_j |Z_-^{1j}|^2 I(m_{\tilde d_{La}}^2 , m_{\tilde e_L}^2 , m_{\chi_j}^2 ) 
\bar u (1+\gamma_5) d \, \bar e (1-\gamma_5) \nu_e ,
\end{eqnarray}
where the mixing matrix elements of the chargino $Z_-^{1j} (j =1,2)$ follow the notation of Rosiek \cite{rosiek}.
The index $j=1,2$ denotes the flavor of the chargino.
We observe that the diagram (i2) has exactly the same couplings, sign and Lorentz structure as (i1).
This fact is consistent with the analysis of the P, CP-odd electron-nucleon interaction at the one-loop level within RPVMSSM of Ref. \cite{yamanaka2}, where similar diagrams with exactly identical RPV couplings appear.
There it was argued that the chargino exchange box diagram is generally smaller than that of the $W$ boson exchange.
This is because the chargino exchange diagram involves three sparticles in the loop.
In this analysis, we will neglect the diagram (i2).

Diagrams (i1) and (i2) are both electroweak corrections to the tree-level contribution (see Fig. \ref{fig:rpv_tree} (t3)).
Due to the flavor change of the $W$ boson, they involve different RPV couplings from the tree-level ones.
This fact provides accessibility to various RPV couplings through beta decay experiments.

Diagrams (i3) and (i4) are suppressed by two factors of light fermion masses, so they can be neglected.

The (i5) and (i6) contributions vanish in the limit of low external and exchanged momenta, so they are also neglected.

\section{\label{sec:obs} Observables in neutron beta decay}

In this section we introduce the observables of the beta decay, i.e. the Fierz interference term and the $D$ correlation.
The general decay distribution of the beta decay is given as follows
\begin{eqnarray}
\omega (E_e , \Omega_e , \Omega_{\nu})&\propto &
1 + a \frac{{\vec p_e} \cdot {\vec p_{\nu}}}{E_e E_{\nu}}
+b \frac{m_e}{E_e} \nonumber \\
&&+ \ \vec \sigma_n \ \cdot 
\left\{
A\frac{{\vec p_e}}{E_e} + B\frac{{\vec p_{\nu}}}{E_{\nu}}
+D\frac{{\vec p_e}\times {\vec p_{\nu}}}{E_e E_{\nu}} 
\right\} \nonumber\\
&&+\ \vec \sigma_e \ \cdot 
\left\{
 N \vec \sigma_n
+ R \frac{ \vec \sigma_n \times \vec p_e}{E_e} 
\right\} \nonumber\\
&&+ \cdots \ .
\label{eq:angulardistribution}
\end{eqnarray}
The Fierz interference term is the shape correction to the beta spectrum ($b \frac{m_e}{E_e}$), and is sensitive to the real part of the scalar interaction of the beta decay ($\bar u d \, \bar e (1-\gamma_5)\nu_e$).
The $D$ correlation is the triple product of the initial neutron spin polarization, emitted neutrino and electron momenta ($D \frac{\vec \sigma_n \cdot \vec p_e \times \vec p_\nu }{E_e E_\nu }$).
It is sensitive to the time reversal violation of the theory, and receives contribution from the imaginary part of the $(V+A)\times (V-A)$ interaction of the beta decay.

Before going to the evaluation of observables, we must first evaluate the nucleon matrix elements of the quark beta decay to derive the effective interaction of the neutron beta decay.
The relevant matrix elements are
\begin{eqnarray}
\langle p | \bar u \gamma^\mu d | n \rangle &=& g_V \, \bar  p \gamma^\mu n , \\
\langle p | \bar u \gamma^\mu \gamma_5 d | n \rangle &=& g_A \, \bar  p \gamma^\mu \gamma_5 n , \\
\langle p | \bar u d | n \rangle &=& g_S \, \bar  p n .
\end{eqnarray}
The vector renormalization constant is $g_V =1 $ by conserved vector current assertion. 
For the axial renormalization constant, we use the experimental value ($g_A = 1.27$).
For the scalar renormalization constant, we take $g_S\approx 0.8$.
We should add some comment on the choice of $g_S$.
Using approximate isospin symmetry, the scalar matrix element can be rewritten as $\langle p | \bar ud | n \rangle = \langle p | \bar uu - \bar dd | p \rangle$.
The latter can be written in terms of the proton-neutron mass splitting $m_p^0 -m_n^0 =-2.05$ MeV (nucleon masses without electromagnetic contribution), up- ($m_u =2.5$ MeV) and down-quark masses ($m_d = 5$ MeV) \cite{pdg} as
\begin{equation}
g_S = \langle p | \bar uu - \bar dd | p \rangle = \frac{m_p^0 -m_n^0 }{m_d - m_u} \approx 0.8\ .
\end{equation}
The up- and down-quark masses are small compared to the typical scale of the QCD, so the above chiral perturbation works well.
We must however note that the input quark masses have a large uncertainty.
The result obtained is consistent with the evaluation of $g_S$ within nonrelativistic quark model \cite{adler}.

From the analysis of the previous section, the one-loop level RPV contribution gives the following beta decay effective interaction:
\begin{equation}
H_{LR} = a_{LR} \cdot \bar p \gamma^\mu (g_V + g_A \gamma_5) n \, \bar e \gamma_\mu (1-\gamma_5) \nu_e +({\rm h.c.}) ,
\label{eq:hlr}
\end{equation}
or
\begin{equation}
H_S = a_S g_S \bar p n \, \bar e (1-\gamma_5) \nu_e +({\rm h.c.}) .
\label{eq:hs}
\end{equation}
From these nucleon level effective interactions, we can derive the Fierz interference term, the $D$ and $R$ correlations of the neutron beta decay as follows \cite{jackson,herczeg2}:
\begin{eqnarray}
D &=& \frac{4g_V g_A }{g_V^2 + 3 g_A^2 } \frac{{\rm Im} (a_{LR}) }{V_{ud} G_F / \sqrt{2}} \ , 
\label{eq:dformula}\\
b &=& 
\frac{2g_V g_S }{g_V^2 + 3 g_A^2 } \frac{{\rm Re} (a_S ) }{V_{ud} G_F / \sqrt{2}} \ , 
\label{eq:b}\\
R &=& 
\frac{-2g_A g_S }{g_V^2 + 3 g_A^2 } \frac{{\rm Im} (a_S ) }{V_{ud} G_F / \sqrt{2}} \ ,
\end{eqnarray}
In this discussion, we will use the Fierz interference term of the analysis of superallowed beta transitions made by Hardy and Towner \cite{hardy}, which provides the most accurate data.
The relation between the Fierz term $b_F$ for the $\beta^+$ decay and the scalar coupling $a_S$ is given as follows:
\begin{equation}
b_F=-\frac{2g_S}{g_V} \frac{{\rm Re}(a_S)}{V_{ud} G_F / \sqrt{2} }\ .
\label{eq:b_f}
\end{equation}
Note that the Fierz term $b_F$ defined above differs from $b$ of eq. (\ref{eq:b}) by a constant factor.
In this discussion, we will use the Fierz term $b_F$ defined in eq. (\ref{eq:b_f}).

There is no tree-level contribution to $a_{LR}$, but at the one-loop level diagrams (a1) and (a2) (see eqs. (\ref{eq:(a1)}) and (\ref{eq:(a2)})) give rise to 
\begin{eqnarray}
a_{LR} &=&  -2 \frac{ \lambda''_{1bi} \lambda''^*_{a1i} V_{ab} m_{u_a} m_{d_b} }{(4\pi)^2 } \frac{G_F}{\sqrt{2}} I(m_{u_a}^2 , m_{d_b}^2 , m_{\tilde d_{Ri}}^2)  \nonumber\\
&&-\frac{ \lambda''_{1bi} \lambda''^*_{a1i} V_{ab} m_{u_a} m_{d_b} A_{u_a} \mu \tan \beta}{(4\pi)^2 }  \frac{G_F}{\sqrt{2}} 
J (m_{\tilde u_L}^2 , m_{\tilde u_R}^2 , m_{\tilde d_L}^2, m_{\tilde d_R}^2) \ .
\label{eq:alr}
\end{eqnarray}
Contribution  of Fig. \ref{fig:loop_wqq} (a2) was also discussed in Ref. \cite{ng}.
Their result is shown to agree with the second term of Eq. (\ref{eq:alr}) thereby noting 
that the definition of the RPV couplings $\lambda''_{ijk}$ and the soft breaking 
term $A_u$ used in Ref. \cite{ng} differs from ours.
By setting $\tan \beta \approx 50$, the sparticle masses, $A_u$ and $\mu$ to 1 TeV, the ratio between 
the first term (contribution of Fig. \ref{fig:loop_wqq} (a1)) and the second term is 
approximately $a_{LR}^{\rm (a1)}/a_{LR}^{\rm (a2)} \approx 0.85$, which gives comparable contributions for both.
The relative sign between them depends on the sign of $A_u$ and cannot be determined by known experimental data, 
so the possibility for both constructive and destructive interferences remains.

On the other hand, there exist both tree and one-loop contributions to $a_S$: $a_S = a_S^{\rm tree}+a_S^{\rm loop}$.
The tree-level effect with selectron ($\tilde e_{Li}$) exchange (Fig. \ref{fig:rpv_tree} (t3)) has been computed in Refs. \cite{herczeg3,yamanaka1} as
\begin{equation}
a_S^{\rm tree}=
-\frac{\lambda_{i11}{\lambda'}_{i11}^*}{4 m_{\tilde e_{Li}}^2} \, .
\label{eq:astree}
\end{equation}
The one-loop diagram (i1) contributes to the Fierz term as (see eq. (\ref{eq:(i1)}))
\begin{equation}
a_S^{\rm loop} = - \frac{\lambda_{i11} \lambda'^*_{ia1}e^2 V_{1a}}{8(4\pi)^2 \sin^2 \theta_W} I(m_{d_a}^2 , m_{\tilde \nu_i}^2 , m_W^2 )\ . \label{eq:as}
\end{equation}
The flavor change due to $W$ exchange gives new contributions with RPV couplings $ \lambda'^*_{i21}$ and $\lambda'^*_{i31}$ in comparison with eq. (\ref{eq:astree}).
Moreover, the sparticle mass dependence is different for $a_S^{\rm tree}$ and $a_S^{\rm loop}$.
These qualitative differences between $a_S^{\rm tree}$ and $a_S^{\rm loop}$ motivate us to consider a particular case in which $a_S^{\rm loop}$ surpasses $a_S^{\rm tree}$.

In later numerical analyses, we do not consider constraint coming from $R$ for the following reason.
It was shown that the experimental data of the EDM of the $^{199}$Hg atom \cite{griffith} can constrain 
the same products of RPV couplings through P, CP-odd electron-nucleon interaction \cite{yamanaka2}.
There the constraints on the combinations of RPV couplings are given at the level of $O(10^{-7})$, which is considerably 
stronger than those which can be given from the present experimental data of the $R$ correlation 
($R_{\rm exp}=0.008\pm 0.011\pm 0.005$) \cite{kozela}.

\section{\label{sec:dcorr}Analysis of the $D$ correlation}

As we mentioned in the introduction, we cannot consider $H_{LR}$ and $H_S$ simultaneously, since baryon number and lepton number violating RPV interactions cannot coexist due to the constraint of the proton lifetime.
We first analyze $H_{LR}$.
As we see in eqs. (\ref{eq:dformula}) and (\ref{eq:alr}),
the $D$ correlation is given in terms of
$\lambda''_{123} \lambda''^*_{112}$, 
$\lambda''_{112} \lambda''^*_{212}$, 
$\lambda''_{123} \lambda''^*_{212}$, 
$\lambda''_{112} \lambda''^*_{312}$, 
$\lambda''_{123} \lambda''^*_{312}$,
$\lambda''_{123} \lambda''^*_{113}$,
$\lambda''_{113} \lambda''^*_{213}$,
$\lambda''_{123} \lambda''^*_{213}$,
$\lambda''_{113} \lambda''^*_{313}$ and
$\lambda''_{123} \lambda''^*_{313}$
(note the antisymmetry in the exchange of the second and third indices for $\lambda''_{ijk}$).
From now on we neglect contributions involving light up- and down-quark masses.
This is also justified by additional suppression due to off-diagonal components of the CKM matrix.
We then obtain the following $D$ correlations for each RPV combination considered:
\begin{eqnarray}
D (\lambda''_{123} \lambda''^*_{212}) &=& 
3.8 \times 10^{-8} \times {\rm Im} (\lambda''_{123} \lambda''^*_{212}) \ , \nonumber\\
D (\lambda''_{123} \lambda''^*_{312}) &=& 
6.3 \times 10^{-5} \times {\rm Im} (\lambda''_{123} \lambda''^*_{312}) \ , \nonumber\\
D (\lambda''_{123} \lambda''^*_{213}) &=&
-2.5 \times 10^{-8} \times {\rm Im} (\lambda''_{123} \lambda''^*_{213}) \ , \nonumber\\
D (\lambda''_{123} \lambda''^*_{313}) &=& 
-6.1 \times 10^{-8} \times {\rm Im} (\lambda''_{123} \lambda''^*_{313})\ ,
\label{eq:dlambda}
\end{eqnarray}
where we have used the quark masses $m_t = 173$ GeV, $m_b=4.2$ GeV, $m_c=1.29$ GeV and $m_s=100$ MeV \cite{pdg}.
The large coefficient for $D (\lambda''_{123} \lambda''^*_{312})$ is due to the large mass of the top quark.
We have set all squark masses to $m_{\tilde q} =1$ TeV (this is the current upper limit given by the LHC \cite{lhc}).
The soft breaking parameters $\mu$ and $A$ have also been set to 1 TeV, and $\tan \beta = 50$.
In this discussion, we have taken the sign of $\mu$ and $A$ so that the contribution from diagrams (a1) and (a2) is constructive.
We must note that the sign of $\mu$ and $A$ parameters are undetermined, and the possibility of cancellation between them exists, as these two contributions have the same order of magnitude.
If the cancellation occurs, the limit on RPV couplings provided by experimental data will be significantly loosened.

Let us consider the possibility to constrain the above combinations of RPV couplings.
The present experimental data for the $D$ correlation are \cite{mumm}
\begin{equation}
D_{\rm exp} = [-0.94\pm 1.89({\rm stat}) \pm 0.97({\rm sys})] \times 10^{-4} \ .
\end{equation}
By comparing with eq. (\ref{eq:dlambda}), we see that the present experimental sensitivity to the $D$ correlation cannot constrain the baryon number violating RPV interactions.
Future experimental progress may however limit them, and we have to discuss it.
The $D$ correlation receives actually an additional contribution from the final state interaction (FSI), of order $O(10^{-5})$ \cite{callan}, and can limit the analysis of the contribution from NP.
Recently, the FSI effect has been evaluated with chiral perturbation to the subleading order \cite{ando}:
\begin{equation}
D_{\rm FSI} = (1.083 \frac{p_e}{p_e^{\rm max}} + 0.228 \frac{p_e^{\rm max}}{p_e} ) \times 10^{-5} \ .
\end{equation}
This provides an accuracy for the $D$ correlation at the percent level.
It is then possible to explore the level of $O(10^{-7})$ for the $D$ correlation with future experiments.
In eq. (\ref{eq:dlambda}), we see that $D$ has a large sensitivity on ${\rm Im} (\lambda''_{123} \lambda''^*_{312})$.
By reaching the sensitivity of $O(10^{-7})$, it will be possible to limit the RPV combination ${\rm Im} (\lambda''_{123} \lambda''^*_{312})$ on the order of $10^{-3}$.
For bilinears $ {\rm Im} (\lambda''_{123} \lambda''^*_{212})$, ${\rm Im} (\lambda''_{123} \lambda''^*_{213})$ and ${\rm Im} (\lambda''_{123} \lambda''^*_{313})$, further experimental developments and theoretical studies to go beyond the $10^{-8}$ sensitivity are needed.

We should also present the constraints given from other analyses for the same baryon number violating RPV interactions.
The first case to consider is the upper limits on single RPV couplings.
Some of the RPV couplings discussed above are actually constrained by the lifetime of the nucleus ($n-\bar n$ oscillation) \cite{barbieri,chang}.
The constraints for RPV couplings relevant in this analysis are
\begin{eqnarray}
|\lambda''_{312}|&<& 2.2 \times 10^{-2} , \nonumber\\
|\lambda''_{313}|&<& 2\times 10^{-2}, 
\end{eqnarray}
where it should be noted that these limits were given by assuming the squark mass $m_{\tilde q} = 200$ GeV.
By respecting the recent lower bounds on squark and gluino masses ($> 1$ TeV) \cite{lhc}, the limits on single RPV interactions should be looser.
Bounds on other single baryon number violating RPV interactions relevant in this analysis have not been worked out yet to our best knowledge \cite{barbier,chemtob}.
We see then that by reaching the sensitivity of $O(10^{-7})$ for the $D$ correlation, it is possible to obtain tighter limits on ${\rm Im} (\lambda''_{123} \lambda''^*_{312})$ than the $n -\bar n$ oscillation data.

The second case to consider is the constraints given by the analysis of the EDM of the neutron and $^{199}$Hg atom.
The bilinears of RPV couplings relevant in our discussion also contribute to the neutron EDM through the EDM of quarks \cite{barbieri} and P, CP-odd 4-quark interaction \cite{ng}.
The quark EDM contribution is estimated to be
\begin{equation}
d_n (d_q) \simeq {\rm Im} (\lambda''^*_{cbi} \lambda''_{a1i}) \frac{e \alpha_{\rm em} V_{ab} V_{c1}}{4\pi^3 \sin^2 \theta_W } \frac{m_{u_a} m_{d_b} m_{u_c} }{ m_W^2 m_{\tilde d_{Ri}}^2} \ ,
\end{equation}
where $a,b$ and $c$ indicate the quark flavor.
The above equation gives the following relations for the products of RPV couplings in our discussion:
\begin{eqnarray}
d_n (d_q ; \lambda''_{123} \lambda''^*_{212}) &\simeq& 4.3 \times 10^{-31} \times {\rm Im} (\lambda''_{123} \lambda''^*_{212}) e\, {\rm cm}\ , \nonumber\\
d_n (d_q ; \lambda''_{123} \lambda''^*_{312}) &\simeq & 1.4 \times 10^{-27} \times {\rm Im} (\lambda''_{123} \lambda''^*_{312}) e\, {\rm cm}\ , \nonumber\\
d_n (d_q ; \lambda''_{123} \lambda''^*_{213}) &\simeq &2.4 \times 10^{-31} \times {\rm Im} (\lambda''_{123} \lambda''^*_{213}) e \, {\rm cm}\ , \nonumber\\
d_n (d_q ; \lambda''_{123} \lambda''^*_{313}) &\simeq & 1.3 \times 10^{-30} \times {\rm Im} (\lambda''_{123} \lambda''^*_{313}) e\, {\rm cm}\ .\nonumber\\
\label{eq:dnlambda}
\end{eqnarray}
For the dependence of the baryon number violating RPV interactions through P, CP-odd 4-quark interaction, relations independent of the model of NP considered can be given as follows (see Ref. \cite{ng} for derivation):
\begin{eqnarray}
|d_n | &=& 1 \times 10^{-19}e\, {\rm cm} \times |D/0.87| \ ,\nonumber\\
|d_{\rm Hg} | &=& 7 \times 10^{-24}e\, {\rm cm} \times |D/0.87| \ .
\end{eqnarray}
These relations can be derived by observing that the right-handed quark current ($\bar u \gamma^\mu (1+\gamma_5) d$) contributing to the $D$ correlation can be coupled to the standard left-handed quark current to form a P, CP-odd 4-quark interaction ($C_{qq}$).
By combining the above formulae with eq. (\ref{eq:dlambda}), we obtain
\begin{eqnarray}
|d_n ( C_{qq};\lambda''_{123} \lambda''^*_{212}) | &=& 
4 \times 10^{-27} e\, {\rm cm} \times {\rm Im} (\lambda''_{123} \lambda''^*_{212}) \ , \nonumber\\
|d_n ( C_{qq};\lambda''_{123} \lambda''^*_{312}) | &=& 
7 \times 10^{-24} e\, {\rm cm} \times {\rm Im} (\lambda''_{123} \lambda''^*_{312}) \ , \nonumber\\
|d_n ( C_{qq};\lambda''_{123} \lambda''^*_{213}) | &=& 
3 \times 10^{-27} e\, {\rm cm} \times {\rm Im} (\lambda''_{123} \lambda''^*_{213}) \ , \nonumber\\
|d_n ( C_{qq};\lambda''_{123} \lambda''^*_{313}) | &=& 
7 \times 10^{-27} e\, {\rm cm} \times {\rm Im} (\lambda''_{123} \lambda''^*_{313})\ . \nonumber\\
\label{eq:dn4qlambda}
\end{eqnarray}
By comparing the above relations with eq. (\ref{eq:dnlambda}), we see that the dependence of RPV interactions on neutron EDM through P, CP-odd 4-quark interaction is much stronger than that given through quark EDM.
Similarly, we obtain the dependence of the EDM of the $^{199}$Hg atom as
\begin{eqnarray}
|d_{\rm Hg} ( C_{qq};\lambda''_{123} \lambda''^*_{212}) | &=& 
3 \times 10^{-31} e\, {\rm cm} \times {\rm Im} (\lambda''_{123} \lambda''^*_{212}) \ , \nonumber\\
|d_{\rm Hg} ( C_{qq};\lambda''_{123} \lambda''^*_{312}) | &=& 
5 \times 10^{-28} e\, {\rm cm} \times {\rm Im} (\lambda''_{123} \lambda''^*_{312}) \ , \nonumber\\
|d_{\rm Hg} ( C_{qq};\lambda''_{123} \lambda''^*_{213}) | &=& 
2 \times 10^{-31} e\, {\rm cm} \times {\rm Im} (\lambda''_{123} \lambda''^*_{213}) \ , \nonumber\\
|d_{\rm Hg} ( C_{qq};\lambda''_{123} \lambda''^*_{313}) | &=& 
5 \times 10^{-31} e\, {\rm cm} \times {\rm Im} (\lambda''_{123} \lambda''^*_{313})\ . \nonumber\\
\label{eq:dhg4qlambda}
\end{eqnarray}
The current experimental data of the neutron EDM are \cite{baker}
\begin{equation}
d_n < 2.9 \times 10^{-26} e\, {\rm cm}.
\label{eq:nedmexp}
\end{equation}
The experimental upper bound of the $^{199}$Hg atom EDM is \cite{griffith}
\begin{equation}
d_{\rm Hg} < 3.1 \times 10^{-29} e\, {\rm cm}.
\label{eq:hgedmexp}
\end{equation}
By combining the above experimental limits with eqs. (\ref{eq:dn4qlambda}) and (\ref{eq:dhg4qlambda}), we see that the current experimental data of the neutron EDM give the tightest constraint on the imaginary part of $\lambda''_{123} \lambda''^*_{312}$ as
\begin{equation}
|{\rm Im} (\lambda''_{123} \lambda''^*_{312})| < 4\times 10^{-3} \ .
\end{equation}
This constraint is tighter than the present experimental limit given by the direct measurement of the $D$ correlation in beta decay, but this argument should be taken with reservation.
We must note that the relation between the neutron EDM and the P, CP-odd 4-quark interaction has a large uncertainty \cite{dn4q}, and there still remain possibilities to have small dependence.
Moreover, the quark EDM contribution to the neutron EDM has also theoretical uncertainty, and the accidental cancellation between it and the P, CP-odd 4-quark interaction effect cannot be completely ruled out.
Therefore it is always of importance to analyze the constraint on RPV interactions which can be derived by the experimental data of the direct measurement of the $D$ correlation.

\section{\label{sec:fierz}Analysis of the Fierz interference term}

Let us now move to the analysis of the Fierz interference term
with lepton number violating RPV interactions.
In this discussion, we consider the one-loop contribution $a_S^{\rm loop}$ as leading, 
while the tree-level $a_S^{\rm tree }$ is small for reasons explained in Section \ref{sec:obs}.
The relevant products of RPV couplings are $\lambda_{i11} \lambda'^*_{ia1}$ (with $i=2,3$ and $a=1,2,3$, see eq. (\ref{eq:as})) and their real part contributes to the Fierz interference term $b_F$.
From the recent update of the analysis on 20 superallowed Fermi transitions, Hardy and Towner have given a new bound on $b_F$ \cite{hardy}.
The result is
\begin{equation}
-\frac{b_F}{2} = +0.0011 \pm 0.0013 \ .
\label{eq:limb_f}
\end{equation}
On the other hand, from eqs. (\ref{eq:as}) and (\ref{eq:b_f}), the RPV contribution to $b_F$ is
\begin{equation}
b_F (\lambda_{i11} \lambda'^*_{ia1}) \approx \frac{2g_S}{(4\pi)^2} \frac{m_W^2}{m_{\tilde \nu_i}^2} \ln \left( \frac{m_{\tilde \nu_i}^2}{m_W^2} \right)  \times \frac{V_{1a}}{V_{ud}} {\rm Re} (\lambda_{i11} \lambda'^*_{ia1}) \ ,
\end{equation}
where $i=2,3$ and $a=2,3$. 
Explicitly, this reads
\begin{eqnarray}
b_F (\lambda_{i11} \lambda'^*_{i21})&=& 7.6 \times 10^{-5} \times {\rm Re} (\lambda_{i11} \lambda'^*_{i21}) \ ,\nonumber\\
b_F (\lambda_{i11} \lambda'^*_{i31})&=& 1.2 \times 10^{-6} \times {\rm Re} (\lambda_{i11} \lambda'^*_{i31}) \ .
\end{eqnarray}
From the above relations and the data of Hardy and Towner (eq. (\ref{eq:limb_f})), we obtain constraints on ${\rm Re } (\lambda_{i11} \lambda'^*_{ia1} )$ as shown in Table \ref{table:rpvlimitstev}.
\begin{table}[htb]
\caption{Upper bounds to the RPV couplings given by the Fierz term of the analysis of Ref. \cite{hardy} for $m_{\tilde \nu_i}=1$ TeV.
Limits from other experiments \cite{chemtob,barbier,kao,rpvphenomenology} are also shown.
}
\begin{ruledtabular}
\begin{tabular}{ccc}
RPV couplings & $b_F$& Other experiments \\ 
\hline
$|{\rm Re } (\lambda_{211} \lambda'^*_{221} )|$& $63$ &$2.9\times 10^{-2}$  \\
$|{\rm Re } (\lambda_{311} \lambda'^*_{321} )|$& $63$ & $1.7\times 10^{-2}$ \\
$|{\rm Re } (\lambda_{211} \lambda'^*_{231} )|$& $4000$ & 0.60 \\
$|{\rm Re } (\lambda_{311} \lambda'^*_{331} )|$& $4000$ & 0.36 \\
\end{tabular}
\end{ruledtabular}
\label{table:rpvlimitstev}
\end{table}
By comparing our result with constraints obtained from other experiments, we see that the upper bounds on RPV couplings from $b_F$ are 4 orders looser for $|{\rm Re } (\lambda_{i11} \lambda'^*_{i21} )|$ and 5 orders for $|{\rm Re } (\lambda_{i11} \lambda'^*_{i31} )|$.
For the Fierz interference term,  there are no FSI contributions \cite{jackson}, so the upper limits on RPV couplings are directly related to the experimental sensitivity.
Future improvement of $b_F$ by more than 4 orders of sensitivity can open accessibility to the combinations of RPV couplings $|{\rm Re } (\lambda_{i11} \lambda'^*_{ia1} )|$ ($i=2,3$ and $a=2,3$).

\section{\label{sec:conclusion}Conclusion}
In this analysis, we have discussed the RPV contribution to the beta decay at the one-loop level.
After careful analysis, we have found that many RPV interactions not relevant at the tree-level contribute at the one-loop level.
For the baryon number violating RPV interactions, combinations
$\lambda''_{123} \lambda''^*_{112}$, 
$\lambda''_{112} \lambda''^*_{212}$, 
$\lambda''_{123} \lambda''^*_{212}$, 
$\lambda''_{112} \lambda''^*_{312}$, 
$\lambda''_{123} \lambda''^*_{312}$,
$\lambda''_{123} \lambda''^*_{113}$,
$\lambda''_{113} \lambda''^*_{213}$,
$\lambda''_{123} \lambda''^*_{213}$,
$\lambda''_{113} \lambda''^*_{313}$ and
$\lambda''_{123} \lambda''^*_{313}$
contribute to the $D$ correlation.
The combination $(\lambda''_{123} \lambda''^*_{312})$ is particularly interesting because by reaching the experimental sensitivity of $O(10^{-7})$ for the neutron beta decay, it is possible to put a $O( 10^{-3})$ constraint on ${\rm Im} (\lambda''_{123} \lambda''^*_{312})$.
This is possible since the FSI contribution to the $D$ correlation is known with the accuracy of $O(10^{-7})$.
The RPV combinations $\lambda''_{123} \lambda''^*_{212}$ and $\lambda''_{123} \lambda''^*_{313}$ can also become interesting because there are no experimental constraints to them so far to our best knowledge.
If the theoretical estimation of the FSI contribution goes beyond the $O(10^{-8})$ level, further experimental progress will give us good chances to probe the corresponding RPV contributions.

For the lepton number violating RPV interactions, the new RPV combinations not relevant at the tree-level $\lambda_{i11} \lambda'^*_{ia1}$ ($i=2,3$ and $a=2,3$) can contribute to the scalar-type interaction of the beta decay through box diagrams involving $W$ boson and chargino.
We have found that the currently known Fierz interference term 
cannot set new limits to Re$(\lambda_{i11} \lambda'^*_{ia1} )$. 
The constraints on Re$(\lambda_{i11} \lambda'^*_{ia1} )$ can however be tightened with further experimental progress by more than 4 orders of sensitivity.

In this analysis, we have seen the importance of the subleading effects.
It has also been emphasized that the access to a variety of RPV interactions through the subleading loop level contributions would be made possible.

%\begin{acknowledgments}
%This work was supported by.....
%\end{acknowledgments}

\appendix

\section{\label{sec:appendix}Flavor conserving electroweak corrections to RPV beta decay}

\begin{figure*}[htb]
\includegraphics[width=16cm]{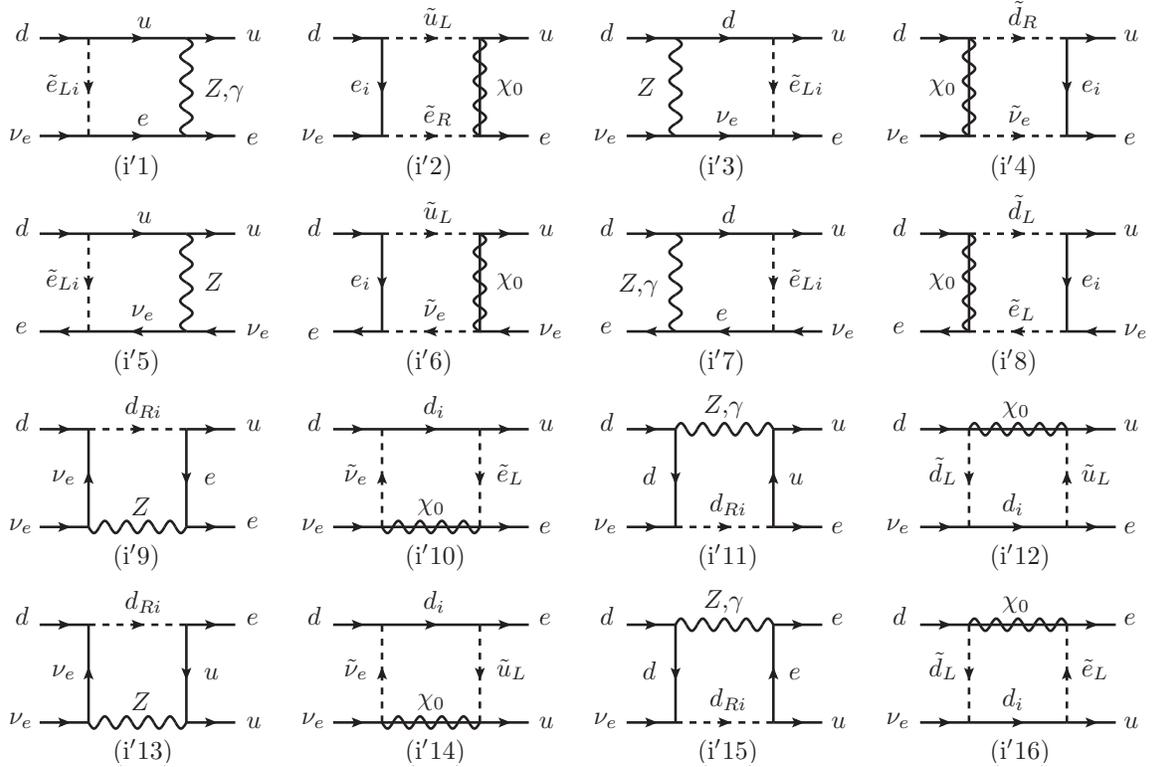}
\caption{\label{fig:fconsbox}
Box diagrams with flavor change contributing to the beta decay at the one-loop level in RPVMSSM.
The neutralino is denoted by $\chi_0$.
}
\end{figure*}

The electroweak flavor conserving corrections (photon, $Z$ boson and neutralino corrections) are seen in detail.
The list of the corresponding diagrams is shown in Fig. \ref{fig:fconsbox}.
Diagrams (i'1) $\sim$ (i'8) are corrections to the tree-level RPV contribution (t3) (see Fig. \ref{fig:rpv_tree}).
These are of scalar-, pseudoscalar-type interaction ($\bar u(1+\gamma_5)d\, \bar e (1-\gamma_5) \nu_e$).
As the $Z$ boson and the photon do not change flavor, the same combination of RPV interactions as the tree-level (t3) ($\lambda_{i11} \lambda'^*_{i11}$, $i=2,3$) is relevant.
Diagrams (i'9) $\sim$ (i'16) are corrections to the tree-level RPV contribution (t4) (see Fig. \ref{fig:rpv_tree}).
These are of $(V-A)\times (V-A)$ type interaction ($\bar u\gamma_\mu (1-\gamma_5)d\, \bar e \gamma^\mu (1-\gamma_5) \nu_e$), and contribute to the shift of the Fermi constant.
They are not interesting in our analysis.


\begin{thebibliography}{99}

\bibitem{herczeg2}
P. Herczeg, Prog. Part. Nucl. Phys. {\bf 46}, 413 (2001).

\bibitem{erler}
J. Erler and M. J. Ramsey-Musolf, Prog. Part. Nucl. Phys. {\bf 54}, 351 (2005).

\bibitem{severijns}
N. Severijns, M. Beck and O. Naviliat-Cuncic, Rev. Mod. Phys. {\bf 78}, 991 (2006).

\bibitem{abele}
H. Abele, Prog. Part. Nucl. Phys. {\bf 60}, 1 (2008).

\bibitem{hewett}
J. L. Hewett {\it et al.}, arXiv:1205.2671 [hep-ex].

\bibitem{jackson}
J. Jackson, S. B. Treiman and H. Wyld, Phys. Rev. {\bf 106}, 517 (1957);
Nucl. Phys. {\bf 4}, 206 (1957).

\bibitem{hardy}
J. C. Hardy and I. S. Towner, Phys. Rev. C {\bf 71}, 055501 (2005); Phys. Rev. C {\bf 79}, 055502 (2009).

\bibitem{halin}
A. L. Hallin, F. P. Calaprice, D. W. MacArthur, L. E. Piilonen, M. B. Schneider and D. F. Schreiber, Phys. Rev. Lett . {\bf 52}, 337 (1984).

\bibitem{soldner}
T. Soldner {\it et al.}, Phys. Lett. B {\bf 581}, 49 (2004);

\bibitem{mumm}
H. P. Mumm {\it et al.}, Phys. Rev. Lett. {\bf 107}, 102301 (2011);
T. E. Chupp {\it et al.}, arXiv:1205.6588 [nucl-ex].




\bibitem{schneider}
M. B. Schneider, F. P. Calaprice, A. L. Hallin, D. W. MacArthur and D. F. Schreiber, Phys. Rev. Lett. {\bf 51}, 1239 (1983).

\bibitem{sromicki}
J. Sromicki {\it et al.}, Phys. Rev. C {\bf 53}, 932 (1996).

\bibitem{kozela}
A. Kozela {\it et al.}, Phys. Rev. Lett. {\bf 102}, 172301 (2009).


\bibitem{herczeg1}
P. Herczeg and I. B. Khriplovich, Phys. Rev. D {\bf 56}, 80 (1997).






\bibitem{mssm}
H. E. Haber and G. L. Kane, Phys. Rept. {\bf 117}, 75 (1985);
J. F. Gunion and H. E. Haber, Nucl. Phys. B {\bf 272}, 1 (1986);
  S.~P.~Martin,
  %``A Supersymmetry primer,''
  arXiv:hep-ph/9709356.
  %%CITATION = HEP-PH/9709356;%%

\bibitem{chemtob}
M. Chemtob, Prog. Part. Nucl. Phys. {\bf 54}, 71 (2005).

\bibitem{barbier}
R. Barbier {\it et al.}, Phys. Rept. {\bf 420}, 1 (2005).

\bibitem{rpvphenomenology}
  G.~Bhattacharyya,
  %``A Brief review of R-parity violating couplings,''
  arXiv:hep-ph/9709395;
  %%CITATION = HEP-PH/9709395;%%
  H.~K.~Dreiner,
  %``An Introduction to explicit R-parity violation,''
  arXiv:hep-ph/9707435.
  %%CITATION = HEP-PH/9707435;%%
  
\bibitem{christova}
E. Christova and M. Fabbrichesi, Phys. Lett. B {\bf 315}, 113 (1993).

\bibitem{drees}
M. Drees and M. Rauch, Eur. Phys. J. C {\bf 29}, 573 (2003).

\bibitem{barger}
V. Barger, G. F. Giudice and T. Han, Phys. Rev. D {\bf 40}, 2987 (1989).

\bibitem{herczeg3}
P. Herczeg, J. Res. Natl. Inst. Stand. Tech. {\bf 110}, 453 (2005).

\bibitem{kao}
  Y.~Kao and T.~Takeuchi,
  %``Single-Coupling Bounds on R-parity violating Supersymmetry, an update,''
  arXiv:0910.4980 [hep-ph].
  %%CITATION = ARXIV:0910.4980;%%

\bibitem{yamanaka1}
N. Yamanaka, T. Sato and T. Kubota, J. Phys. G {\bf 37}, 055104 (2010).

\bibitem{ng}
J. Ng and S. Tulin, Phys. Rev. D {\bf 85} 033001 (2012).

\bibitem{yamanaka2}
N. Yamanaka, Phys. Rev. D {\bf 85}, 115012 (2012).
%arXiv:1204.6466 [hep-ph].

\bibitem{smirnov}
A. Yu. Smirnov and F. Vissani, Phys. Lett. B {\bf 380}, 317 (1996);
G. Bhattacharyya and P. B. Pal, Phys. Lett. B {\bf 439}, 81 (1998); Phys. Rev. D {\bf 59}, 097701 (1999).

\bibitem{rosiek}
J. Rosiek, Phys. Rev. D {\bf 41}, 3464 (1990).
%\url{http://www.fuw.edu.pl/~rosiek/physics/prd41.html}.


\bibitem{pdg}
K. Nakamura {\it et al.} (Particle Data Group), J. of Phys. G {\bf 37}, 075021 (2010) and 2011 partial update for the 2012 edition (URL: \url{http://pdg.lbl.gov});
J. Beringer {\it et al.} (Particle Data Group), Phys. Rev. D {\bf 86}, 010001 (2012).

\bibitem{adler}
S. Adler {\it et al.}, Phys. Rev. D {\bf 11}, 3309 (1975).

\bibitem{griffith}
W. C. Griffith, M. D. Swallows, T. H. Loftus, M. V. Romalis, B. R. Heckel and E. N. Fortson, Phys. Rev. Lett. {\bf 102}, 101601 (2009).

\bibitem{lhc}
ATLAS Collaboration (Georges Aad {\it et al.}), Phys. Rev. Lett. {\bf 106}, 131802 (2011); Phys. Lett. B {\bf 701}, 186 (2011); 
  %``Searches for supersymmetry with the ATLAS detector using final states with
  %two leptons and missing transverse momentum in sqrt{s} = 7 TeV proton-proton
  %collisions,''
%  arXiv:1110.6189 [hep-ex];
Phys. Lett. B {\bf 709}, 137 (2012);
  %%CITATION = ARXIV:1110.6189;%%
CMS Collaboration (Vardan Khachatryan {\it et al.}), Phys. Lett. B {\bf 698}, 196 (2011); 
  %``Searches for supersymmetry in final states with leptons or photons and
  %missing energy,''
%  arXiv:1111.2733 [hep-ex].
Proc. Sci. {\bf EPS-HEP2011}, 271 (2011).
  %%CITATION = ARXIV:1111.2733;%%






\bibitem{callan}
C. G. Callan and S. B. Treiman, Phys. Rev. {\bf 162}, 1494 (1967).

\bibitem{ando}
S. Ando, J. McGovern and T. Sato, Phys. Lett. B {\bf 677}, 109 (2009).


\bibitem{barbieri}
R. Barbieri and A. Masiero, Nucl. Phys. B {\bf 267}, 679 (1986).

\bibitem{chang}
D. Chang and W.-Y. Keung, Phys. Lett. B {\bf 389}, 294 (1996).

\bibitem{baker}
C. A. Baker {\it et al.}, Phys. Rev. Lett. {\bf 97}, 131801 (2006).

\bibitem{dn4q}
V. M. Khatsimovsky, I. B. Khriplovich and A. S. Yelkhovsky, Ann. Phys. {\bf 186}, 1 (1988);
D. Demir, O. Lebedev, K. A. Olive, M. Pospelov and A. Ritz, Nucl. Phys. B {\bf 680}, 339 (2004);
H. An, X. Ji, F. Xu, JHEP {\bf 1002}, 043 (2010);
E. Mereghetti, J. de Vries, W. H. Hockings, C. M. Maekawa and U. van Kolck, Phys. Lett. B {\bf 696}, 97 (2011).



\end{thebibliography}
\end{document}